# Graphene-based perfect optical absorbers harnessing guided mode resonances

M. Grande,[1,*] M. A. Vincenti,[2] T. Stomeo,[3] G. V. Bianco,[4] D. de Ceglia,[2] N. Aközbek,[5]

V. Petruzzelli,[1] G. Bruno,[4] M. De Vittorio,[3,6] M. Scalora[7] and A. D'Orazio[1]

[1] *Dipartimento di Ingegneria Elettrica e dell'Informazione, Politecnico di Bari,*
*Via Re David 200, 70125 Bari, Italy*
[2] *National Research Council, Charles M. Bowden Research Center, RDECOM,*
*Redstone Arsenal, Alabama 35898-5000 – USA*
[3] *Center for Bio-Molecular Nanotechnologies, Istituto Italiano di Tecnologia (IIT),*
*Via Barsanti, 73010 Arnesano (Lecce), Italy*
[4] *Istituto di Nanotecnologia, Nanotec - CNR*
*Via Orabona 4, 70126 Bari, Italy*
[5] *AEgis Technologies Inc,*
*410 Jan Davis Dr, Huntsville - AL, 35806*
[6] *Dipartimento Ingegneria dell'Innovazione, Università Del Salento,*
*Via Arnesano, 73100 Lecce, Italy*
[7] *Charles M. Bowden Research Center, RDECOM,*
*Redstone Arsenal, Alabama 35898-5000 – USA*
[*]*marco.grande@poliba.it*

**Abstract**

We numerically and experimentally investigate graphene-based optical absorbers that exploit guided mode resonances (GMRs) achieving perfect absorption over a bandwidth of few nanometers (over the visible and near-infrared ranges) with a 40-fold increase of the monolayer graphene absorption. We analyze the influence of the geometrical parameters on the absorption rate and the angular response for oblique incidence. Finally, we experimentally verify the theoretical predictions in a one-dimensional, dielectric grating and placing it near either a metallic or a dielectric mirror.

## 1. Introduction

In the last years, two-dimensional materials have attracted a great interest due to their unprecedented properties [1]. The experimental isolation of the first two-dimensional material, graphene, has accelerated different research lines in this new field, paving the way for new photonic applications and functionalities [2].

In this framework, devices devoted to light absorption played an important role thanks to high monolayer graphene absorption that reaches a constant value of about 2.3% over the visible range [2], well greater than the absorption values of metallic and dielectric materials having same thickness. However, when a perfect absorption (or total light absorption, TLA) is required, several approaches have been tackled to achieve 100% absorption. A first solution relates to the coherent perfect absorption (CPA) where a lossy slab, illuminated by both sides, is able to absorb all the incoming light since it acts as an "anti-laser" [3,4,5]. In this case, the amplitude and the phase of both the beams must be controlled in order to achieve coherent absorption (or an absorptive interferometer). Furthermore, the material absorption does not need to be necessarily high, therefore it may show a low single-pass absorption rate (e.g. wavelength range close to the transparent range). For example, Reference [3] details how a 100 µm-thick silicon slab, working at ~945 nm (where the silicon extinction coefficient $k$ is extremely low, i.e., $8 \cdot 10^{-4}$) is able to totally absorb the incoming light at a specific wavelength (making the device narrow-band).

This system is equivalent to a two-port optical system and the CPA phenomenon is related to the poles and zeros (i.e. singularities) of its (*non hermitian*) scattering matrix $S$ with respect to the input and the output fields. When CPA constrains are satisfied, the combination of interference and dissipation (lossy slab) leads to perfect absorption. This approach overcomes the maximum achievable absorption, equal to 50%, when only a single input beam is considered in symmetric configurations. Finally, it is worth noting that this phenomenon does not require highly resonant structure ($Q$ factor is about 840 in [4]) or surface nano-structuring. On the contrary, it requires very thick slabs, if the slab thickness (100 micron) and the working wavelength (about 900 nm) are compared, and necessitates a complete control on the intensity and phase of the incoming beams in the two-port system. A similar approach has been recently proposed in reference [6] for thin meta-materials but, in this case, a plasmonic thin film has been patterned in order to satisfy the CPA conditions.

Up to now, CPA in graphene has been efficiently achieved only in the THz regime [7-9], since its optical parameters and its thickness do not satisfy CPA conditions in the visible range.

However, it is possible to reduce the complexity of the CPA systems to one-port systems by adding a perfect mirror to the two-port system: intuitively, in this case, the perfect mirror reflects the transmitted light, which in turn, generates the incoming beam at the second port. In this one-port system, the absorption $A$ is equal to $A=1-R$, where $R$ is the reflectance and the transmission channel *T* is suppressed (*T*=0). Therefore, also in this one-port system, interference and absorption can again be "mixed" and designed to achieve a null reflectance that leads to perfect absorption or total light absorption.

One example corresponds to the "Salisbury screen" (SS) in the microwave regime [10] that is essentially constituted by a mirror, a lossless spacer and a thin absorbing layer. In particular, a fraction of the input beam that is transmitted into the spacer undergoes multiple partial reflections by resonating between the absorbing layer/spacer interface and the mirror. The total reflected beam can be designed to destructively interfere attaining total absorption of the incoming light. Recently, different schemes, based on the SS configuration, have been proposed in the optical regime [11-14]: in particular, in configurations based on graphene, the bi-dimensional layer plays only the role of the thin absorbing material (non-patterned graphene) [12] or adds a resonant structure due to its plasmonic nature (patterned graphene) [13-14].

Since all these (CPA and SS) configurations define a cavity (e.g. an asymmetric Fabry-Pérot cavity with two mirrors having different reflectivities), they can also be analyzed in the domain of critical coupling [15,16] as reported in [17] where a very thin absorbing film is placed on a dielectric Bragg reflector. However, the main drawback of the configurations reported in [12, 17] is that optical Salisbury screen operates efficiently only at one wavelength value (at which destructive interference occurs) and, consequently, it cannot be tuned unless the refractive index, or alternatively the spacer thickness, is changed. On the contrary, periodic structures allow phase matching and hence a tuning of the operating wavelength.

Finally, when the graphene is employed as absorbing film in the visible range, the lossless spacer and the dielectric mirror do not suffice to achieve perfect absorption as demonstrated in [18]. A solution can be found by adding a second mirror as demonstrated in [19] where the graphene is sandwiched between two dielectric mirrors: the first one is a partially reflecting mirror whilst the second mirror acts as a perfect mirror with high reflectivity (it can be considered almost 100%). The presence of the cavity gives rise to an enhancement of the electric field that strongly interacts with the absorbing layer leading to a resonant cavity-enhanced absorption.

In conclusion, we can say that all these results clearly show that is possible to employ graphene to achieve perfect absorption even if it shows a low single-pass absorption in different spectral ranges.

In this scenario, we propose to study optical absorbers that exploit guided mode resonances (GMRs) [20-21], also known as Fano resonances [22-23], in order to attain perfect absorption in the near-infrared (NIR) range. In particular, we combine a resonant structure, i.e. a dielectric one-dimensional (1D) grating, with a reflection mirror separated by a lossless spacer. It is interesting to point out that the dielectric 1D grating based absorber without the mirror can achieve a theoretical maximum absorption of about 60% as demonstrated in our recent work [24]. Furthermore, we consider the influence of the geometrical parameters of the dielectric 1D grating on the absorbing behavior and its angular response when the impinging source is tilted. Finally, we experimentally verify the theoretical assumption by realizing the dielectric 1D grating and placing it on two different mirrors, a metallic and a dielectric one, respectively.

## 2. Numerical results

Figure 1(a) shows the sketch of the proposed configuration that is equivalent to a one-port system as described in the introduction. The mirror reflects the transmitted light, suppressing the transmission channel and generating a reflected beam towards the resonant structure.

Figure 1(b) details the geometrical parameters of the dielectric 1D grating based absorber made of polymethyl-methacrylate (PMMA) stripes deposited on a tantalum pentaoxide ($Ta_2O_5$) slab that is supported by a silicon dioxide ($SiO_2$) substrate. The monolayer graphene is sandwiched between the polymeric layer and the $Ta_2O_5$ slab forcing it to interact with the guided mode resonances. The $Ta_2O_5$ slab thickness $t_{Ta2O5}$, the periodicity $p$, the PMMA width $w_{PMMA}$ and the PMMA thickness $t_{PMMA}$ are initially set equal to 100 nm, 470 nm, 305 nm ($w_{PMMA}$=0.65$p$) and 650 nm, respectively.

Finally, a semi-infinite flat mirror (either an ideal mirror or gold), that reflects the transmitted light, supports the 1D dielectric grating.

Simulations are based on the Rigorous Coupled-Wave Analysis (RCWA) method. The refractive index of the dispersive materials (*PMMA*, *Ta₂O₅*, *SiO₂* and *Au*) was retrieved by means of ellipsometric measurements. We found that the permittivity values do not differ from the data reported in [25] and the extinction coefficients for all the dielectric media (except for gold and the monolayer graphene) are almost zero allowing us to deal with a lossless system. The monolayer graphene was described following the model reported in [26]. Hereinafter, we will only consider TE polarization (plane of incidence *yz* and electric field along *x*), although similar results and considerations have been obtained for TM polarization.

We start our analysis by considering the dielectric 1D grating sketched in Figure 1 supported by a perfect (ideal) mirror: in particular, we introduce (i) a perfect magnetic mirror, PMM (or magnetic wall) and (ii) a perfect electric mirror, PEM (or perfect conductor). Figures 2(a) and (b) show the absorption map and absorption maxima, in presence of monolayer graphene, in the wavelength range of interest (0.6 μm – 0.8 μm) for the $SiO_2$ thickness $t_{SiO2}$ varied in the range 0 – 0.4 μm at normal incidence. It is evident that the absorption wavelength shifts towards higher wavelengths as the $SiO_2$ layer thickness is increased. Furthermore, the absorption Full-Width at Half-Maximum (FWHM) is about 5 nm for both mirrors, when the Fano feature associated with the TE guided mode resonance is considered, and it is almost constant for all the $SiO_2$ thicknesses.

At the same time, the absorption varies between about 10% and perfect absorption: in particular, the PMM configuration (Figure 2(b)) shows three maxima (located at $t_{SiO2}$ =0.105 μm, 0.190 μm and 0.355 μm) and two minima (located at $t_{SiO2}$=0.140 μm and 0.390 μm) while the PEM configuration (Figure 2(d)) shows two maxima (located at $t_{SiO2}$=0.230 μm and 0.315 μm) and one minimum (located at $t_{SiO2}$=0.265 μm).

All the maxima correspond to a 43-fold increase of the isolated monolayer graphene absorption (2.3%) since the monolayer graphene is the only absorbing material in both the configurations.

It is also possible to observe a shift of maxima and minima of the PMM case with respect to the PEM configuration that is related to the phase difference introduced by the two boundary conditions: in fact, the reflection coefficients are equal to 1 and -1 in the case of PMM and PEM configuration, respectively. This difference introduces a π phase change (i.e. the PEM mirror reverses the phase with respect to the PMM mirror) while the thickness delta $\Delta_t$, equal to about 0.125 μm, introduces another π phase change corresponding to about $2kn_{SiO2}\Delta_t$ (the $SiO_2$ refractive index is almost constant over the range of interest) leading to a total 2π phase difference. Finally, it is possible to verify that the distance between maxima (minima) (e.g. in Figure 2(b)) corresponds to $\lambda/2n_{SiO2}$ due to constructive (destructive) interference of the Fabry-Pérot cavity mode in the $SiO_2$ layer.

Figures 3(a) and (b) show the absorption map and absorption maxima, respectively, in the wavelength range of interest (0.6 µm – 0.8 µm), in presence of the monolayer graphene for each SiO$_2$ thickness $t_{SiO2}$ in the range 0 – 0.3 µm, at normal incidence, when a real gold mirror is considered. Also in this case, Figures 3(a) and (b) show a similar behavior as in Figures 2(c)-(d) with a shift of about 30 nm that is related to the complex reflection coefficient (due to the complex gold refractive index) that introduces a phase shift. Moreover, this configuration allows to achieve a perfect absorption and this behavior can be easily explained by inspecting the magnetic field profile ($H_z$ field component) at the maximum ($t_{SiO2}$ = 0.3 µm) that reveals the localization of the field at the monolayer graphene position ($y$ = 0.1 µm) as illustrated in Figure 3(c). It is worth noticing that the magnetic field slightly penetrates into the metal layer ($y$ < -0.3 µm) due to its finite conductivity. On the contrary, the magnetic field profile of the configuration corresponding to the minimum of Figure 3(b) shows that the field is localized in the SiO$_2$ spacer scarcely interacting with the monolayer graphene (Figure 3(d)). Lastly, if we keep constant the wavelength, i.e. λ=0.7416 µm, the absorption varies between 8.5% and 100% (Figure 3(a)).

The performance of the dielectric 1D grating based absorber, at normal incidence, when the PMMA width $w_{PMMA}$ is varied is reported in Figure 4. The spacer thickness is fixed equal to $t_{SiO2}$ = 0.3 µm (Figure 3(b)) so the perfect absorption condition is fulfilled. In particular, Figure 4(a) shows that absorption wavelength (0.737 µm at $w_{PMMA}$ = *0.5p*, spanning in the range 0.735 µm - 0. 740 µm) and its FWHM (4.9 nm at $w_{PMMA}$ = *0.5p*) are almost constant in the $w_{PMMA}$ range 0.3*p* – 0.7*p*. This is also valid for the absorption that shows a maximum equal to 100% when $w_{PMMA}$ = 0.65*p* while it reaches 97.6% in the center (i.e. $w_{PMMA}$ = 0.5*p*) as evidenced in Figure 4(b). This behavior confirms the stability of the proposed device against the fabrication tolerances and proves a stable critical coupling over a wide range of the geometrical parameters. It is worth reminding that the leakage is proportional to the fill factor that in this case linearly varies with the PMMA width $w_{PMMA}$.

Up to now, we only considered the response of the optical absorbers at normal incidence. Since the proposed device is essentially based on a grating configuration, we are also interested in its angular behavior. Figure 5(a) shows the absorption map when the incident angle θ (Fig. 1(a)) is varied in the range 0° - 20°. From the inspection of the plot, it is evident that the resonance at normal incidence splits in two arms when the source is tilted. Furthermore, the position of the two absorbing peaks follows the dispersion of the guided mode resonance that introduces the rapid variation of the intensity in the spectrum: in particular, since we are in the zero-order diffraction regime (at input interface) due to the sub-wavelength period, the two arms follow the coupling, into the $Ta_2O_5$ slab, with the evanescent mode orders having *m* equal to +/-1. Figure 5(b) shows the dispersion relation for the GMR analytically calculated by considering the diffraction behavior of the four-layers structure.

Finally, it is interesting to note that (i) the main feature of Figure 5(a) is related to the asymmetric spectrum of the two arms that show different diffraction efficiencies and (ii) even if the proposed device is sensitive to incident source angle, one can reduce the angular dependence by working in a wavelength region where multiple anti-crossing points occur: for example, in the range 5°-20°, the absorption is limited between the wavelength range 0.65 μm and 0.675 μm, i.e. over a narrower bandwidth.

## 3. Fabrication and experimental results

In order to validate the theoretical results shown in the previous section, we initially fabricated the dielectric 1D grating on a 500 μm thick $SiO_2$ substrate. In particular, a 100 nm-thick $Ta_2O_5$ slab was grown on this substrate by means of a sputtering system while a Chemical Vapour Deposition (CVD) monolayer graphene was manually transferred onto the $Ta_2O_5$ slab. Then, the PMMA stripes, with a width $w_{PMMA}$ and a periodicity *p* equal to 0.65*p* and 470 nm, were realized by means of an electron beam lithography system (Raith150) operating at 20 kV with an area dose equal to 300 μC/cm$^2$.

Figure 6 shows the Raman spectrum of the transferred monolayer graphene, after the fabrication process, proving the almost absence of defects (i.e. D peak at about 1350 cm$^{-1}$) and, hence, the preservation of the graphene quality during the fabrication process. The inset in Figure 6 portrays the Scanning Electron Microscope (SEM) image of the 1D grating composed by the periodic PMMA stripes.

The configuration sketched in Figure 1(a) could be realized by covering the $SiO_2$ substrate back-side with a metal layer. However, we decided to separate the mirror from the dielectric 1D grating in order to use the same sample with different mirrors. In particular, we first placed the dielectric 1D grating supported by the $SiO_2$ substrate on a 0.1μm-thick gold layer deposited on a silicon wafer.

We experimentally characterized the complete structure by means of an ad-hoc optical setup. In particular, we shined white light on the sample by means of a 5x microscope objective, with a numerical aperture NA equal to 0.1. An optical iris spatially filtered the impinging light in order to only select the patterned area. Then, the reflected light was sent to an optical spectrometer (HR4000 from Ocean Optics) through collecting system composed by a multimode optical fiber and an aspherical lens.

Figure 7 shows the experimental (blue solid curves) reflectance (labeled with $R$) and absorption (labeled with $A$, where $A = 1-R$) spectra of the composed device that reveal an asymmetric dip and an asymmetric peak, respectively. This behavior is fully consistent with the asymmetry depicted in Figure 5(a). In the same Figure, we also superimposed the numerical results (red dashed curves) showing a very good agreement with the experimental counterparts.

Finally, as mentioned above, the fabrication of the dielectric 1D grating on a 500 μm thick $SiO_2$ substrate allowed us to change easily the reflecting mirror. In this regard, we also placed the dielectric 1D grating on a commercial broadband dielectric mirror (BB05-E02 supplied by Thorlabs).

Figures 8(a) and (b) show the numerical reflectance and the absorption spectra of the composed device for both the broadband dielectric mirror (solid lines) and gold layer (dashed lines). The comparison reveals that the two mirrors show almost an identical response and, hence, validates the possibility to interchange different dielectric or metallic reflecting mirrors. This behavior can be explained by firstly considering that the thick substrate prevents the change of the effective refractive index of the dielectric 1D grating by the presence of the mirror; furthermore, the mirrors only introduce a phase shift that is averaged.

Finally, figures 8(c) and (d) show the experimental reflectance and the absorption spectra of the same device for both the broadband dielectric mirror (solid lines) and gold layer (dashed lines). In the same plots, we also considered two different structures, one with lower dose (250 μC/cm$^2$, red curves) and one with higher dose factor (350 μC/cm$^2$, green curves) revealing a slight shift with respect to the simulated one (300 μC/cm$^2$, blue curves). The comparison between Figures 8(a) and (b) and the blue curves in Figure 8(c) and (d) show a very good agreement.

## 4. Conclusion

In conclusion, we have detailed the numerical analysis and experimental verification of an optical absorber that exploits GMRs in order to attain perfect absorption. In particular, our numerical findings reveal that the use of a lossless spacer and an ideal mirror is sufficient to achieve perfect absorption with a 40-fold increase of the monolayer graphene absorption over a bandwidth of ~5 nm. Similar behavior is found when a gold mirror replaces the ideal mirror. In this case, the minimum whole thickness of the optical absorber is equal to only about 1 μm, highlighting the compactness of the device. At the same time, for a fixed wavelength (e.g. $\lambda$ = 0.7416 μm), it could be possible to control the absorption rate from about 10% to 100% tuning the spacer thickness. In this direction, one could envision the possibility to tune the spacer refractive index exploiting the metallic mirror and the monolayer graphene as contacts.

Then, the angular dependence analysis reveals an asymmetric absorption response that follows the guided mode resonance due to the presence of the PMMA grating. Furthermore, even if this is beyond the scope of this work, the angular dependence could be "reduced" by exploiting the multiple anti-crossing regions in the diffraction curves, leading to almost angular insensitive optical absorbers.

From an experimental point of view, we fabricated the dielectric 1D grating and placed it on two different mirrors, a metallic (gold) one and a dielectric one achieving a very good agreement between experimental results and numerical simulations. We have also shown that this approach permits to interchange mirrors and this could also be used to tune and switch between two different absorbing regimes with and without mirror.

The enormous advantage of the proposed geometry resides also in the possibility to directly deposit or transfer the monolayer graphene on a flat surface and not on a nanostructured surface significantly reducing the issues of the fabrication process.

All these features clearly demonstrate that the proposed approach can be exploited to efficiently design and optimize tunable graphene-based optical absorbers.

**Acknowledgements**

M. Grande thanks the U.S. Army International Technology Center Atlantic for financial support (W911NF-13-1-0434). This research was performed while the authors M. A. Vincenti and D. de Ceglia held National Research Council Research Associateship awards at the U. S. Army Aviation and Missile Research Development and Engineering Center.

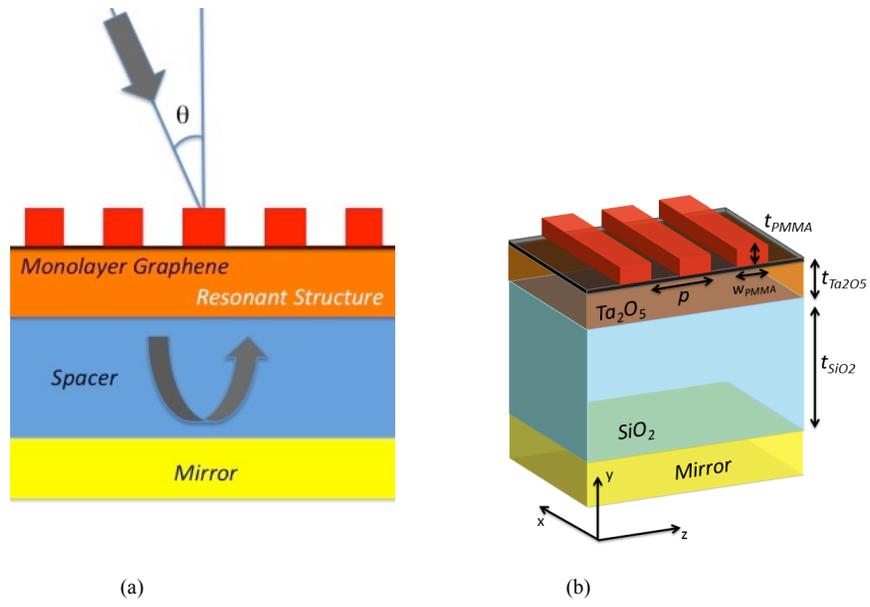

Fig. 1. (a) One-port system with a resonant structure, a spacer and a reflecting mirror. (b) Sketch of the 1D grating: PMMA stripes (red) on $Ta_2O_5$ slab (orange) grown on silicon dioxide substrate (cyan). The black thin layer indicates the monolayer graphene.

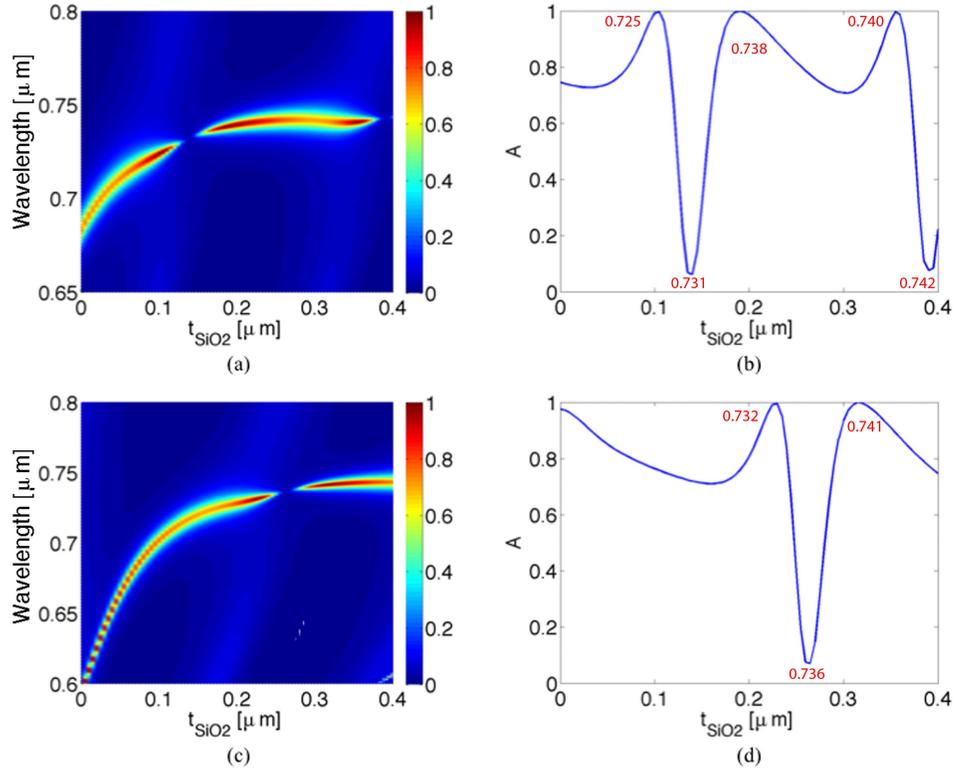

*Fig. 2. Absorption map versus the $t_{SiO2}$ thickness for the (a) PMM and the (c) PEM configurations, in presence of the monolayer graphene, at normal incidence. Maximum achievable absorption when the $t_{SiO2}$ thickness is varied for (b) PMM and (d) PEM configurations when PMMA width $w_{PMMA}$ is set equal to 0.65p. The maxima and minima wavelengths (in micron) are indicated in red.*

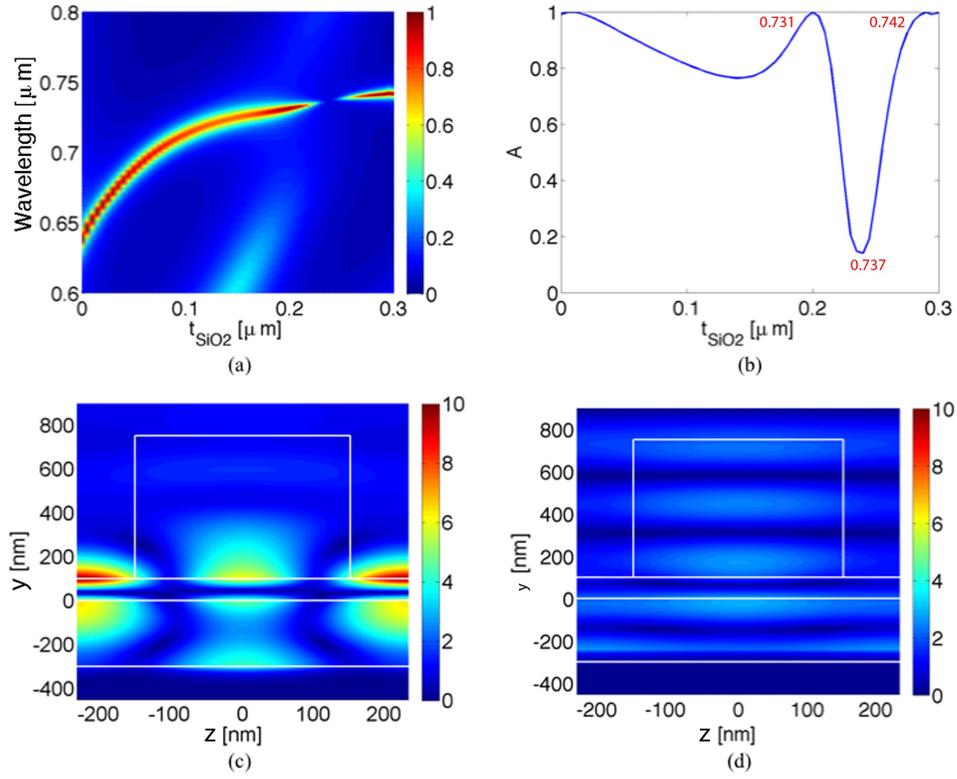

*Fig. 3. (a) Absorption map versus the $t_{SiO2}$ thickness and (b) its maximum at normal incidence when PMMA width $w_{PMMA}$ is set equal to 0.65p. The maxima and minima wavelengths (in micron) are indicated in red. Magnetic field profile (amplitude of the $H_z$ component) when $t_{SiO2}$ is equal to (c) 0.3 μm at λ = 0.7418 μm and (d) 0.24 μm at λ = 0.7368 μm, respectively. The monolayer graphene is positioned at y = 0.1 μm (white horizontal line) while the gold layer is placed at y < -0.3 μm.*

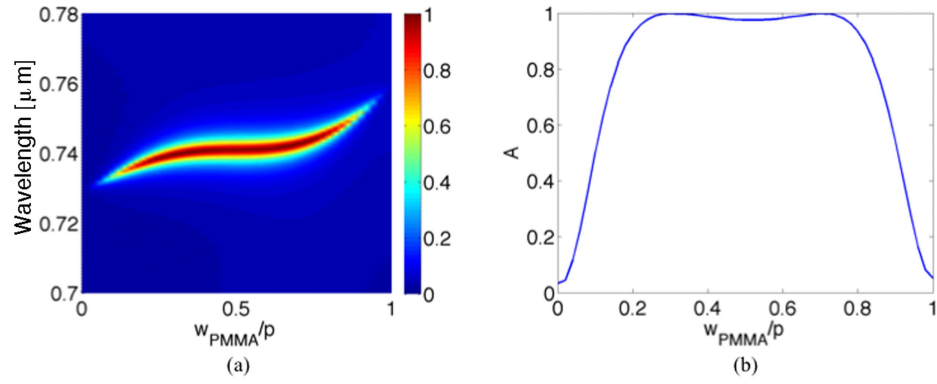

*Fig. 4. (a) Absorption map and (b) maximum achievable absorption versus the PMMA width $w_{PMMA}$ normalized to the periodicity p when $t_{SiO2}$ = 0.3 μm and the monolayer graphene is taken into account.*

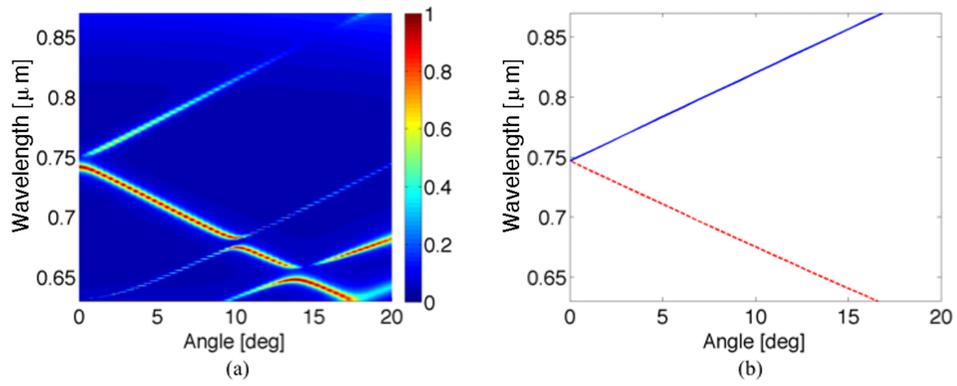

*Fig. 5. (a) Angular dependence of the absorption when the incident angle θ is varied in the range 0° - 20°; (b) Guided resonance mode wavelength dispersion curve versus the angle of incidence for m=1 (blue solid curve) and m=-1 (red dashed curve), respectively.*

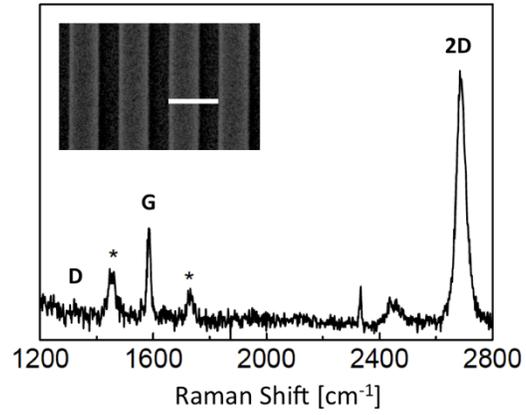

*Fig. 6. Raman spectrum of the monolayer graphene after the fabrication process; the asterisks refer to the PMMA stripe spectral features; (inset) Scanning Electron Microscope (SEM) micrograph of the fabricated device where the white scalebar refers to 470 nm.*

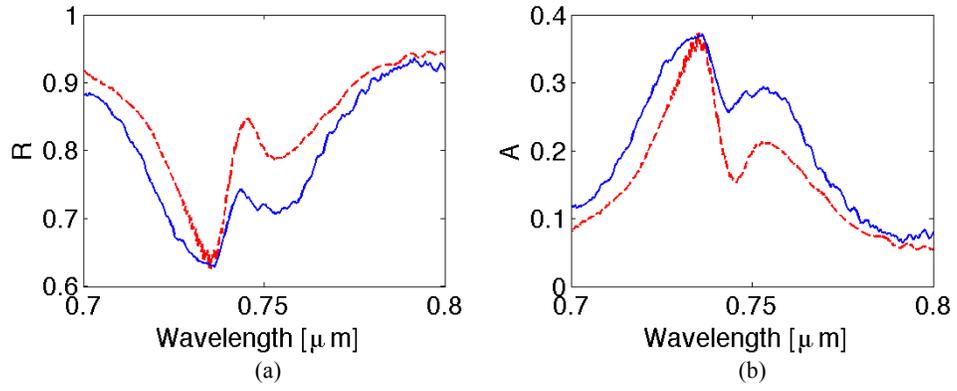

*Fig. 7. Experimental (a) reflectance and (b) absorption spectra of the fabricated device (blue solid curves). For comparison, the red dashed curves refer to the numerical response.*

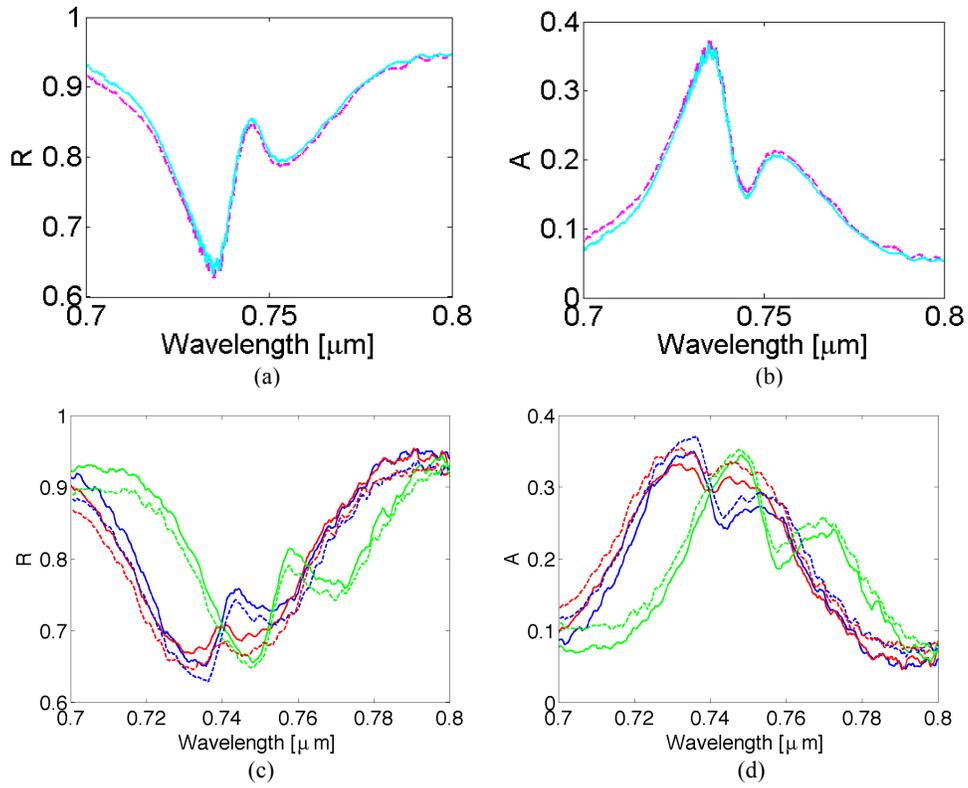

*Fig. 8. Simulated (a) reflectance and (b) absorption spectra for the dielectric mirror (solid lines) and for the gold mirror (dashed lines), respectively. Measured (c) reflectance and (d) absorption spectra when the dose is equal to 250 µC/cm$^2$ (red curve), 300 µC/cm$^2$ (blue curve) and 350 µC/cm$^2$ (green curve) for the dielectric mirror (solid lines) and for the gold mirror (dashed lines), respectively.*